\documentclass{iau}
\usepackage{graphicx}
\usepackage{gensymb}

\title[Swift follow-up of GW170817 and improvements for the future] 
{GW170817: \textit{Swift} UV detection of a blue kilonova, and improving the search in O3}

\author[Aaron Tohuvavohu \& Jamie A. Kennea]   
{Aaron Tohuvavohu$^1$, Jamie A. Kennea$^2$ \and the \textit{Swift} gravitational wave follow-up group}

\affiliation{Department of Astronomy and Astrophysics, The Pennsylvania State University, University~Park, PA 16802, USA\\ email: {\tt $^1$ aaronb@swift.psu.edu $^2$ kennea@swift.psu.edu}}

\pubyear{2017}
\volume{338}  
\pagerange{119--126}
\jname{Gravitational Wave Astrophysics: Early results from GW searches and electromagnetic counterparts}
\editors{G. Gonzalez ed.}
\begin{document}

\maketitle

\begin{abstract}
\textit{Swift}'s rapid slewing, flexible planning, and multi-wavelength instruments make it the most capable space-based follow-up engine for finding poorly localized sources. During O1 and O2 \textit{Swift} successfully tiled hundreds of square-degrees of sky in the LVC localization regions, searching for, and identifying, possible X-ray and UV/O transients in the field. \textit{Swift} made important contributions to the discovery and characterization of the kilonova AT~2017gfo, discovering the UV emission and providing the deepest X-ray upper limits in the first 24 hours after the trigger, strongly constraining the dynamics and geometry of the counterpart. \textit{Swift} tiled 92\% of the galaxy convolved error region down to average X-ray flux sensitivities of $10^{-12}$ erg $\mbox{cm}^{-2}$ $\mbox{s}^{-1}$, significantly increasing our confidence that AT 2017gfo is indeed the counterpart to GW 170817 and sGRB 170817. However, there remains significant room for improvement of \textit{Swift's} follow-up in preparation for O3. This will take the form of both revised observation strategy based on detailed analysis of the results from O2, and significant changes to \textit{Swift}'s operational capabilities. These improvements are necessary both for maximizing the likelihood that \textit{Swift} finds a counterpart, and minimizing the impact that follow-up activities have on other \textit{Swift} science priorities. We outline areas of improvement to the observing strategy itself for optimal tiling of the LVC localization regions. We also discuss ongoing work on operational upgrades that will decrease latency in our response time, and minimize impact on pre-planned observations, while maintaining spacecraft health and safety.
\end{abstract}

\firstsection 
\section{Introduction}
The NASA \textit{Swift} Mission (Gehrels et al. 2004) is a mid-sized Explorer class satellite designed to study Gamma-Ray Bursts (GRBs). \textit{Swift} has partnered with the LIGO-VIRGO collaboration (LVC) in performing a follow-up program to search for electromagnetic (EM) counterparts to gravitational wave (GW) events since the 2009-2010 science run (\cite{Evans 2012}). The general scheme is a tiling campaign that uses \textit{Swift}'s two narrow field-of-view instruments, the X-ray Telescope (XRT) and UV/Optical Telescope (UVOT), to target high probability fields within the gravitational wave localization region. The \textit{Swift} spacecraft is unique in its ability to carry out tiling campaigns of this type, because of its fast-response and extremely flexible scheduling and commanding, and its singularly rapid slewing. This tiling capability was developed with LIGO tiling in mind, but was extensively tested both by observing simulated LIGO triggers from O1, but also, starting in June 2016, for the \textit{Swift} SMC Survey (S-CUBED; Kennea et al., \textit{in prep.}), a weekly survey covering the Small Magellanic Cloud in 142 pointings of 60 seconds each. Lessons learned in safely performing these rapid re-pointing maneuvers with the spacecraft, building the complex schedules to facilitate the campaigns, and understanding the necessary instrument modes to be able to perform useful science in such brief snapshots were developed by the \textit{Swift} team and brought to bear on the goal of rapidly searching the gravitational wave localization regions. During the post-detection era (O2) \textit{Swift} has responded to many LVC triggers in this capacity, searching hundreds of square-degrees of galaxy convolved probability for a possible EM counterpart using the techniques described by Evans et al (2016). 

Because all of the LVC triggers in O1, and most of those in O2, were BBH events, the follow-up campaigns that \textit{Swift} performed put strong limits on the X-ray emission from these events, and were consistent with the expectation that BBH mergers would be dark, but were most useful as test runs to ensure that \textit{Swift}'s response would be optimal when a gravitational wave signal from a system with an expected EM counterpart finally was detected. Results from these campaigns were used to help improve the tiling strategy itself, reduce the response time of the observatory, and test the automated data processing and transient identification pipelines. 

\section{\textit{Swift}'s response to GW/sGRB 170817}
At 12:41:04 UT the LVC registered a gravitational wave signal whose strain waveforms were consistent with a binary neutron star system (\cite{Abbot}). Approximately 2 seconds later \textit{Fermi}'s Gamma-Ray Burst Monitor (GBM) triggered on a short ($\sim$ 2 seconds duration) gamma-ray signal consistent with the GW localization. At this time, \textit{Swift} was at a location in its orbit such that the GRB and GW localization regions were occulted by the Earth limb (Figure 1), and so the relevant region of the sky was not visible to \textit{Swift}'s Burst Alert Telescope (BAT).

\begin{figure}[h]
\centering
\includegraphics[width=.75\textwidth]{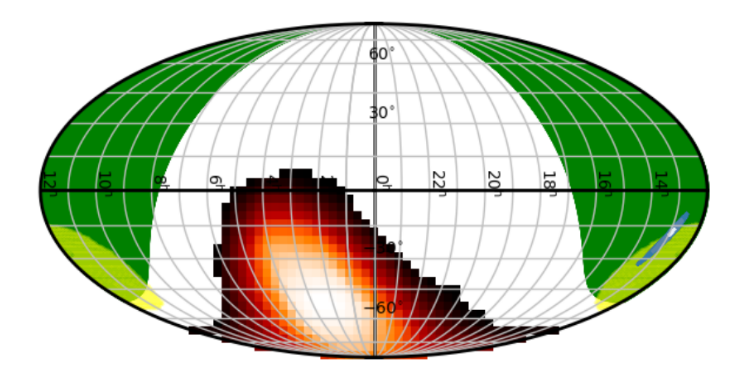}
\caption{The BAT FoV at the time of the GW/sGRB trigger. The black/red/white region shows the BAT FoV, with the color scale indicating the partial coding fraction. Everything occulted by the Earth from the perspective of \textit{Swift} is shaded green. The blue and yellow regions are the GW and \textit{Fermi} GBM 90\% confidence localization regions respectively.}
\end{figure}

News of a GW signal temporally coincident with a \textit{Fermi} GBM trigger was disseminated to the community via a private GCN circular at 13:21 UT (\cite{epoch}). Upon receipt, the \textit{Swift} team immediately triggered a follow-up campaign. \textit{Swift}'s narrow-field instruments began performing settled observations in search of an EM counterpart at 13:37 UT, 16 minutes after notification. \textit{Swift} was the first telescope to begin pointed observations in response to the trigger. At this time, the best localization available came from GBM (90\% containment area of $\sim$ 1626 $\mbox{deg}^2$) as the multiple detector LIGO sky map took several hours to produce because of a glitch in the LIGO-Livingston data. For this reason, \textit{Swift}'s earliest observations were a 37-point tiling covering the $\sim 3.7$ $\mbox{deg}^2$ at the center of the \textit{Fermi} GBM localization region (see Figure 2).

\begin{figure}[h]
\centering
\includegraphics[width=.75\textwidth]{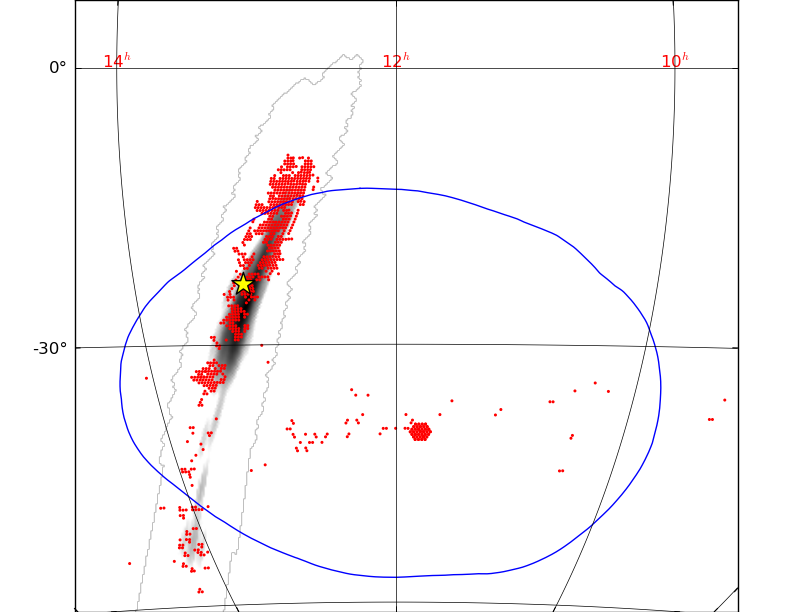}
\caption{Sky map of \textit{Swift} observations in equatorial coordinates. Each red pixel is a single XRT observation of 120~s. The dark grey region shows the extent of the final, 3-detector GW localization. The location of AT 2017gfo is marked with a yellow star. The early 37-point tiling can be seen at the center of the blue \textit{Fermi} GBM $90\%$ confidence region. The widely scattered fields were from the first iteration of the observing plan, which was based on the single-detector GW sky map available at the time. }
\end{figure}

At 17:54 UT a three-detector sky map with a 90\% containment area of 33.6$\mbox{deg}^2$ became available from the LVC. This sky map was convolved with the GWGC catalog, as described in \cite{Evans 2016}. \textit{Swift} promptly began a tiling campaign targeting these galaxies, with 120~s of exposure per tile. In this campaign \textit{Swift} observed 92\% of the galaxy convolved localization region over the course of 744 individual pointings, finding no evidence for new bright X-ray transients in the field down to a flux upper limit of $f_\mathrm{X}\geq10^{-12}$ erg cm$^{-2}$ s$^{-1}$. Many UV/optical transients were found but none were deemed plausible counterpart candidates. 

In order to determine useful constraints from \textit{Swift}'s non-detection in the wide-field search a simulation of 10,000 sGRB afterglows based on \textit{Swift}'s flux-limited sample (\cite{davanzo}) was performed. These light curves were rescaled to the distance derived from the GW signal, and positions were drawn at random from the 3-D `bayestar-HLV' sky map (\cite{map}). We then determined which of the simulated events were in galaxies observed by XRT as part of our tiling campaign, and if the source flux would have been detectable by \textit{Swift} at the time of the observation. It was found that in 65\% of cases the source would have been detected. However, sGRB 170817 was quite under-luminous compared to the \textit{Swift} sample. Re-scaling the predicted X-ray flux for the simulated sGRB afterglows down by a factor of $10^3$, we find that were there such an afterglow \textit{Swift} would have detected it 11\% of the time.

\begin{figure}[h]
\centering
\includegraphics[width=.75\textwidth]{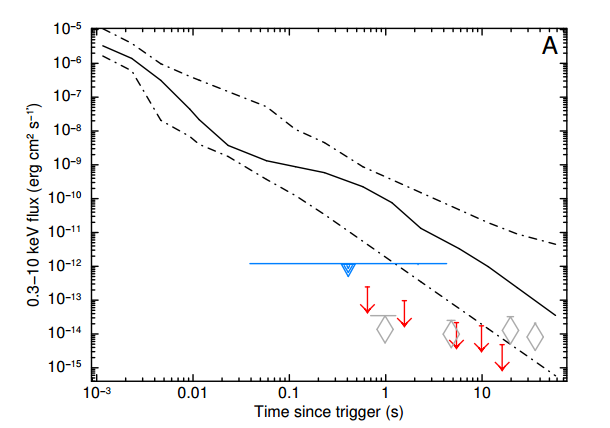}
\caption{The median behavior and 25-75\% distribution of sGRB afterglows, scaled to $40$ Mpc, are shown by the solid and dashed lines respectively. The blue line with the triangle shows the time-span over which the tiling observations were taken, and the typical flux sensitivity achieved per tile. The red arrows and grey diamonds show X-ray flux upper limits on emission from AT 2017gfo from \textit{Swift} and NuSTAR observations respectively.}
\end{figure}

At 01:05 UT on August 18, 2017 the One-Meter Two-Hemisphere (1M2H) team reported the detection of a possible counterpart associated with the galaxy NGC 4993 (\cite{Coulter}), later designated AT 2017gfo. At the time, the nature of this candidate, and whether it was indeed connected to GW/sGRB 170817, was unclear. \textit{Swift} continued its wide-field tiling campaign, but regularly interrupted the tiling to slew to the field of NGC 4993, with the first observation beginning $\sim 2$ hours after the announcement.

\section{Observations of AT 2017gfo}
\begin{figure}[h!]
\centering
\includegraphics[width=\textwidth]{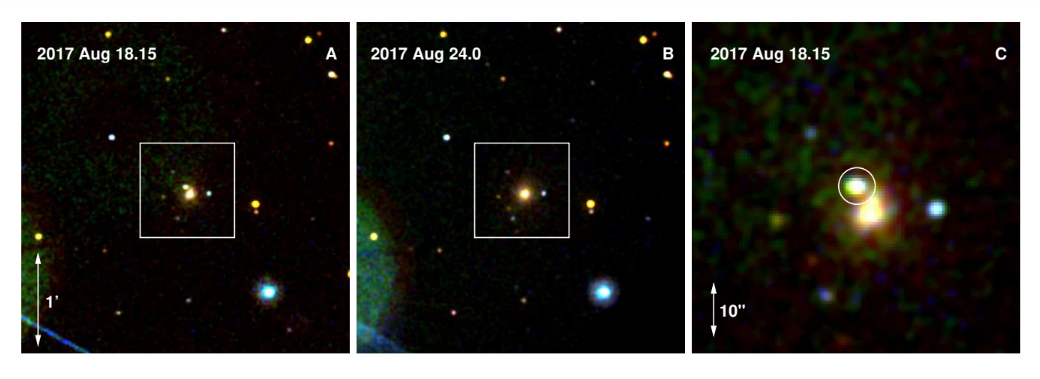}
\caption{\textit{Swift} UVOT image of the field of NGC 4993. Panel A shows the first epoch, in which bright UV emission is clearly seen.  Panel B demonstrates the rapid fading at blue wavelengths. Panel C shows the first epoch zoomed in with AT 2017gfo circled. The \textit{u, uvw1, uvm2} filters are assigned to the red, green and blue channels respectively.}
\end{figure}

\textit{Swift} performed the first observation of AT 2017gfo at 03:34 UT on August 18, 2017. In the first exposure \textit{Swift} detected the UV emission from this source, and provided the earliest sensitive X-ray upper limits. AT 2017gfo was detected as a bright and fading UV source with an initial magnitude of $u=18.9^{+0.09}_{-0.08}$ mag (AB). Figure 5 shows the \textit{Swift} UV and optical light curves, plotted alongside optical and near-infrared photometry from Pan-STARRS (\cite{panstarss}). The rapid fading at UV wavelengths is clearly evident, in contrast to the optical and near-infrared emission which remained detectable for over a week (\cite{kasliwal}). No X-ray source was found in this first observation with a 3-$\sigma$ (0.3 - 10 keV) upper limit of $5.5\times10^{-3}$ cts $\mbox{s}^{-1}$. Assuming an energy conversion typical for GRB afterglows and the distance to NGC 4993 of $\sim40$ Mpc, this gives $L_x < 4.2 \times 10^{40}$~erg~$\mbox{s}^{-1}$ at 0.6 days after the merger. \textit{Swift} continued monitoring this source with both XRT and UVOT over the 15 days between discovery and when the source became too close to the Sun to observe on September 2, 2017. In this time \textit{Swift} collected 172 Ks of observations, in which no X-ray source was detected. Summing all of the X-ray data gives a 3-$\sigma$ upper limit of $2.8 \times 10^{-4}$ cts $\mbox{s}^{-1}$. With the same assumptions as above this gives $L_x< 2.1\times 10^{39}$ ergs $\mbox{s}^{-1}.$ 

\begin{figure}[h]
\centering
\includegraphics[width=.75\textwidth]{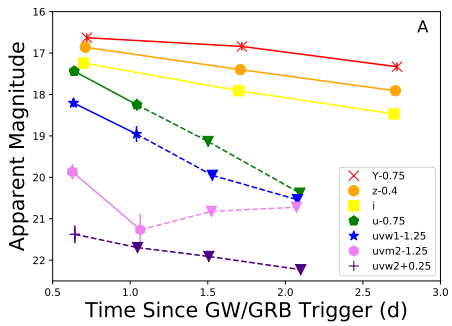}
\caption{\textit{Swift} UVOT light curve of AT 2017gfo. Upper limits are shown by inverted triangles. The \textit{i,z,y} bands are from Pan-STARRS and are included for the purpose of comparison. All data are corrected for host galaxy contamination.}
\end{figure}
The absence of X-rays at early times, as well as the early UV emission allow independent constraints on the nature of the counterpart. As shown in Figure 3, the lack of X-rays from AT 2017gfo make its interpretation as an on-axis sGRB afterglow a tenuous one. While there are a few sGRBs in the sample with afterglows that fade extremely quickly, and \textit{Swift} would therefore have missed, the vast majority of the sGRB afterglow population would have been detectable. Using the analytic afterglow formulation for synchrotron emission (\cite{Evans2017}), we can derive physical constraints from the X-ray upper limits. For on-axis viewing the X-ray non-detection independently limits the amount of energy coupled to relativistic ejecta to less than $10^{50}$~erg, assuming isotropic emission. The X-ray observations can also help inform our understanding of the origin of the UV emission. The optical-to-X-ray spectral index $\beta_{OX}\geq1.6$ at 0.6~days after the merger is far outside the norm for GRB afterglows (\cite{Horst}). 

The UV/optical SED further discredits the interpretation of the UV counterpart as the emission from a typical sGRB afterglow. The SED can be fit well with a blackbody function with $T_{BB}=7300\pm200$~K at 0.6 days after the merger, with rapid cooling to $T_{BB}=6400\pm200$~K at 1.0~days after the merger. The blackbody model yields an average fit statistic of $\chi^2=2.2$ (5 degrees of freedom). For comparison, if the origin of the UV counterpart is from synchrotron afterglow emission, one would expect a power-law model of the form $f_\lambda\propto\lambda^{-\alpha}$ to describe the SED well. At both 0.6 and 1.0 days after the merger we find $\alpha\sim1.0$. However, the fit quality is extremely poor, with an average fit statistic of $\chi^2=98.15$ (5 degrees of freedom). As such, the UV emission must have its origin in a different physical process than an on-axis GRB afterglow. We find that an off-axis orientation of $\approx 30\degree$ fits the early absence of X-rays along with the late-time Chandra detection well (see Figure 6). Likewise, the presence of UV emission independently constrains the orientation from the other end as sight lines in the plane of the merger are expected to have significantly reduced UV emission due to the presence of lanthanide-rich dynamical ejecta. Further implications of the UV emission as well as modeling that shows broad consistency between the early UVOT observations and kilonova models can be found in \cite{Evans2017}.
\begin{figure}[h]
\centering
\includegraphics[width=.69\textwidth]{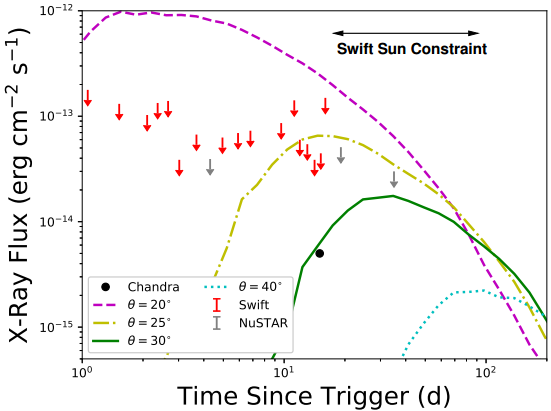}
\caption{Simulated X-ray afterglow lightcurves for a typical GRB, at different viewing angles. Here $E_{AG}=2\times10^{51}$ erg, $n_0=5\times10^{-3}$ $\mbox{cm}^{-3}$ and $\theta_{jet}=$0.2 radians. }
\end{figure}
\firstsection
\section{Future Improvements}
While \textit{Swift}'s contributions to the discovery of AT 2017gfo were important and largely successful, there remains room for improvement for future LIGO/Virgo runs such as the upcoming O3. \textit{Swift} is the only satellite observatory capable of performing a fast-response targeted tiling campaign, and will remain alone in this role for O3. In addition, depending on the time at which the GW event is detected, \textit{Swift}'s 96~min orbit means that fields become observable in $<1$~hour from detection, whereas ground-based telescopes may have to wait many hours for optimal observing conditions. For this reason the \textit{Swift} team is investing considerable effort into improving response time, tiling strategy, transient detection pipelines, scheduling, and further study into maintaining spacecraft health and safety while performing these campaigns.
\subsection{Scheduling}
\textit{Swift} has the most complex and demanding scheduling of any satellite observatory, performing on average $\sim120$ individual observations per day (compared to a handful per day for most other missions) while juggling a host of astronautical and scientific constraints. Over the lifetime of the \textit{Swift} mission various planning and scheduling tools have been developed to automate as much of this task as possible. However, the construction of optimal schedules has remained the domain of human Science Planners, who outperform the automated tools by significant margins. Unfortunately, it takes a Science Planner approximately 5 hours, on average, to construct an optimal observing plan for one 24-hour period.  For rapid response campaigns this is simply too slow. Because of this, the current automated pipeline that constructs the rapid response tiling campaigns, and associated schedules, produces plans that are scientifically sub-optimal. 
In order to rapidly construct these schedules, the LIGO tiling observations are merged with current pre-planned observations for the day, bumping those observations that conflict with LIGO tile observations. In order to fix any constraint violating issues with the plan that result from this merger, further science targets are dropped in favor of a target from a large pool of `fill-in' targets, typically targets for which observations are not time critical, to prioritize the health and safety of \textit{Swift}.

What this means is that even those pre-approved targets that have locations on the sky that are not at all in conflict with the GW localization region are often bumped out of the schedule, and the spacecraft spends a significant amount of time observing targets that are not of the highest scientific merit. While this was deemed tolerable during O1 and O2, given the expected increased rate of GW triggers in O3 consistently dropping \textit{Swift}'s pre-planned science becomes unacceptable. As such, significant effort has been invested in building a fast, effective, and optimal automated planning pipeline that can outperform humans. This project is described in \cite{aaron} and will have the added benefits of increasing the efficiency of the observatory in general as well as significantly adding to its already prodigious fast-response capabilities.
\subsection{Commanding}
The most important limiting factor with respect to improving the satellite's response time is the commanding delay. While individual pointings and pre-loaded tiling patterns, like the 37-point tiling that covered the center of the \textit{Fermi} GBM localization region, can be uploaded through the Tracking and Data Relay Satellite System (TDRSS) with a 25~minute delay at any time of the day, the more complex custom tiling to cover the GW error regions currently require a ground station pass. The time between these passes vary, but if one is unlucky and receives a trigger notification right after a pass, there may be a multiple hour wait before the next one. In the case of GW 170817, a pass begin almost immediately after receiving the epochal GCN (\cite{epoch}), which is what allowed for the rapid 16 minute response time.  The \textit{Swift} team is currently in the late stages of testing a system, called PPTOO (Pre-Planned Target of Opportunity), that will effectively allow for any set of commands (including tiling campaigns for custom GW error regions) to be uploaded through the TDRSS system. This will ensure a 25~minute response time even in the worst case scenario of a multiple-hour wait until the next ground-station pass.

\subsection{Tiling Strategy} 
\textit{Swift} began its tiling campaign based on the 3-detector sky map at $\sim$ 18:00 UT, many hours before the sun set in Chile, meaning \textit{Swift} was able to be on-target before most of the ground-based telescopes capable of observing the location of GW 170817. The host galaxy of AT 2017gfo, NGC 4993, was one of the most massive, and therefore most probable, galaxies in the field. Indeed NGC 4993 was in the first 10 galaxies observed by the ground observatories. Given the fact that \textit{Swift} began observing hours earlier and spent at most 120 s per field, why did \textit{Swift} not observe AT 2017gfo first?

The reason lies in the fact that \textit{Swift} did not observe the galaxies in strict probability order. The tiling campaigns were constructed by optimizing the schedule on minimum slew distance between tiles as well as probability. This choice was made based on studies performed with the 2-detector sky maps from events in O1 and the beginning of O2. The localization regions from these events were quite large, were often bimodal, and had regions that were spatially disconnected. Large sets of simulations were performed to determine the best way to tile these regions. It was found that the probability of finding a counterpart was actually increased if the list was weighted by slew distance in addition to probability. Essentially, in the case of spatially disconnected probability regions the spacecraft can spend large amounts of time slewing back-and-forth across the sky, to the point that the observing-to-slew time ratio drops significantly enough such that the spacecraft can actually observe more cumulative probability in the same amount of time if one weights the list by slew distance.

However, GW170817 had a very small localization region. Therefore, had \textit{Swift} simply worked its way down the galaxy list in strict probability order it would have observed NGC 4993, and hence AT 2017gfo, at 18:09 UT on August 17, 5 hours before the discovery by ground telescopes. Given the knowledge that \textit{Swift} can find these counterparts, and potentially provide extremely important observations at very early times, it is worthwhile to improve our response to ensure that we succeed in this capacity in O3.

\section{Conclusion}
\textit{Swift}'s fast and wide response provided the only sensitive UV and X-ray transient search in the localization region, and increased the confidence that AT 2017gfo was in fact the counterpart to GW and sGRB 170817. Moreover, \textit{Swift}'s early discovery of UV emission from AT 2017gfo and the absence of early X-rays provide independent constraints on the dynamics and orientation of the kilonova. \textit{Swift} is a unique facility and the only satellite observatory capable of carrying out these rapid response custom tiling campaigns. \textit{Swift} will continue its GW follow-up activities in O3 with significant improvements in many facets of its operations including commanding latency, scheduling, and tiling strategy. These operational improvements will also significantly benefit other multi-messenger and localization projects, and further ensure that \textit{Swift} is able to maximally contribute to the new age of gravitational wave and multi-messenger astronomy.

\end{document}